\documentclass[apj,twocolumn]{oja}

\usepackage[T1]{fontenc}
\usepackage{aas_macros}
\usepackage{natbib}
\usepackage{times}

\usepackage{amsmath}
\usepackage{booktabs}
\usepackage{multirow}
\usepackage{color}
\usepackage{soul}
\usepackage[breaklinks,colorlinks,citecolor=blue,urlcolor=blue]{hyperref}
\usepackage{orcidlink}
\usepackage{xspace}

\newcommand{\package}[1]{\textsl{#1}}

\newcommand{\be}{\begin{equation}}
\newcommand{\ee}{\end{equation}}
\newcommand{\bea}{\begin{eqnarray}}
\newcommand{\eea}{\end{eqnarray}}

\newcommand{\Gyr}{\,\mathrm{Gyr}} 
\newcommand{\MSun}{\,\mathrm{M_{\odot}}} 
\newcommand{\mgal}{M_{\star, \mathrm{gal}}}
\received{2024 December 6}
\revised{X{\small xxxx} XX}
\accepted{X{\small xxxx} XX}

\shorttitle{The cosmic GC formation history}
\shortauthors{Joschko et al.}

\begin{document}\sloppy\sloppypar\raggedbottom\frenchspacing

\title{The cosmic globular cluster formation history in the E-MOSAICS simulations\vspace{-15mm}}

\author{Philipp S.~Joschko$^{1,2}$}
\author{J.~M.~Diederik~Kruijssen$^\star$\,\orcidlink{0000-0002-8804-0212}$^{3,4}$}
\author{Sebastian Trujillo-Gomez\,\orcidlink{0000-0003-2482-0049}$^{5}$}
\author{Joel~L.~Pfeffer\,\orcidlink{0000-0003-3786-8818}$^{6}$}
\author{\\Nate Bastian\,\orcidlink{0000-0001-5679-4215}$^{7,8}$}
\author{Robert~A.~Crain\,\orcidlink{0000-0001-6258-0344}$^{9}$}
\author{Marta Reina-Campos\,\orcidlink{0000-0002-8556-4280}$^{10,11}$}
\thanks{$^\star$E-mail: \href{mailto:kruijssen [at] coolresearch.io}{kruijssen [at] coolresearch.io}}

\affiliation{$^1$Astronomisches Rechen-Institut, Zentrum f\"{u}r Astronomie der Universit\"{a}t Heidelberg, M\"{o}nchhofstra\ss{}e 12-14, 69120 Heidelberg, Germany}
\affiliation{$^2$Institute for Theoretical Physics, Heidelberg University, Philosophenweg 12, 69120 Heidelberg, Germany}
\affiliation{$^3$Cosmic Origins Of Life (COOL) Research DAO, \href{https://coolresearch.io}{https://coolresearch.io}}
\affiliation{$^4$Technical University of Munich, School of Engineering and Design, Department of Aerospace and Geodesy, Chair of Remote Sensing Technology, Arcisstr.~21, 80333 Munich, Germany}
\affiliation{$^5$Astroinformatics Group, Heidelberg Institute for Theoretical Studies, Schloss-Wolfsbrunnenweg 35, 69118 Heidelberg, Germany}
\affiliation{$^6$Centre for Astrophysics \& Supercomputing, Swinburne University, Hawthorn, VIC 3122, Australia}
\affiliation{$^7$Donostia International Physics Center (DIPC), Paseo Manuel de Lardizabal, 4, E-20018 Donostia-San Sebasti\'{a}n, Guipuzkoa, Spain}
\affiliation{$^8$IKERBASQUE, Basque Foundation for Science, E-48013 Bilbao, Spain}
\affiliation{$^9$Astrophysics Research Institute, Liverpool John Moores University, 146 Brownlow Hill, Liverpool L3 5RF, UK}
\affiliation{$^{10}$Department of Physics \& Astronomy, McMaster University, 1280 Main Street West, Hamilton, L8S 4M1, Canada}
\affiliation{$^{11}$Canadian Institute for Theoretical Astrophysics (CITA), University of Toronto, 60 St George St, Toronto, M5S 3H8, Canada\\}


\begin{abstract}\noindent
We present a comprehensive analysis of globular cluster (GC) formation and evolution across the $34^3$~Mpc$^3$ volume of the E-MOSAICS galaxy formation simulations. Defining GCs as surviving, high-mass ($>10^5~\MSun$) clusters, we analyse their formation histories as a function of their metallicity and host galaxy mass, also distinguishing between central and satellite galaxies. The redshift of peak GC formation rate increases weakly with galaxy mass, decreases with metallicity, and does not differ between centrals and satellites. The epoch of peak GC formation precedes that of the stars by a factor of $1.1{-}1.6$, primarily due to `downsizing', i.e.\ low-mass galaxies form their stars later. Consequently, this offset decreases with galaxy mass, leading to nearly coeval stellar and GC populations in massive galaxies ($>10^{11}\MSun$). GCs themselves do not exhibit strong downsizing, because they predominantly formed at early cosmic epochs conducive to the formation (through high gas pressures) and survival (through high galaxy merger and GC migration rates) of massive, compact stellar systems. The total GC formation rate in the volume peaks at $z\approx 2.5$, shortly before star formation peaks at $z\approx 2$, but well after the general cluster formation rate at $z\approx 4$, reflecting a survivor bias where surviving GCs formed more recently. We find that GC formation commenced early, at $z>10$, such that the results of this work may provide a framework for interpreting direct observations of proto-GC formation with the \textit{JWST}, especially as these observations accumulate to enable statistical studies.\\
\end{abstract}

\section{Introduction}
\label{sec:introduction}

Globular clusters (GCs) are thought to be among the oldest stellar populations in the Universe \citep[e.g.][]{harris91,brodie06,forbes18}. Because GC properties have been shaped by the galactic environment across cosmic history \citep[e.g.][]{elmegreen10,kruijssen15b}, their old ages make GCs useful probes of the early galaxy formation process \citep[e.g.][]{Choksi_2018,ChoksiGnedin_2019,Kruijssen_2019a,kruijssen20c,trujillogomez21}. To probe early galaxy formation with GCs, it is necessary to either measure GC ages with the precision needed to determine exactly at which epoch of galaxy formation GCs formed, or to take observations at the sensitivity required to directly detect proto-GCs forming at high redshift \citep[e.g.][]{adamo20}.

In recent years, major progress has been made on both fronts. Age determinations are pushing towards ever-increasing precision, enabling GC ages to be measured also in galaxies beyond the Milky Way \citep[e.g.][]{usher19,cabreraziri22}. Additionally, the arrival of the \textit{James Webb Space Telescope} (JWST) has allowed observations to routinely observe proto-GCs in their nascent environments at high redshift \citep[e.g.][]{mowla22,vanzella22,claeyssens23}. For the first time, these developments enable a comprehensive assessment of when GCs formed across cosmic history, as a function of the host galaxy properties. By integrating over the galaxy population, these efforts will eventually enable the construction of a `Madau plot' \citep[in original form describing the cosmic star formation history]{Madau_2014} showing the cosmic GC formation history.

Attempts to uncover the cosmic GC formation history can be informed only by a limited range of priors. Within the Milky Way, GCs are found across a broad range of ages between 6~Gyr and a Hubble time \citep[e.g.][]{marinfranch09,dotter10,dotter11,forbes10,vandenberg13,Kruijssen_2019b}. However, the Milky Way is only a single galaxy, and differences in galaxy formation histories are expected to lead to differences in GC age distributions \citep[e.g.][]{Kruijssen_2019a}. Pioneering observations of the GC age distribution in extragalactic systems confirm this variety of formation histories \citep[e.g.][]{Parisi+2014,Beasley+2015,usher19,Garro+2021}. This variance has the important implication that observations and predictions of the cosmic GC formation history both require integrating over as wide a range of galaxies as possible.

In this paper, we aim to provide model predictions for the cosmic GC formation history using a suite of hydrodynamical cosmological simulations from the E-MOSAICS project \citep{Pfeffer_2018,Kruijssen_2019a}. These simulations span a cubic cosmic volume of 34.3~Mpc in side length, sufficient to probe the variety of galaxies needed to capture the formation histories of most GCs in the Universe \citep{harris16}. With this study, we extend our previous work predicting the GC formation histories of Milky Way-like galaxies \citep{Reina-Campos_2019,keller20}, with the specific goal of providing a robust modelling framework for predictions in the era of direct proto-GC observations with JWST.

The structure of this paper is as follows. In Section~\ref{sec:E-MOSAICS}, we summarise the simulation suite used in this work. In Section~\ref{sec:results}, we present and discuss the predicted GC formation histories and compare these with those of stars and all stellar clusters. We also show how the typical GC formation age depends on galaxy mass, GC metallicity, and galaxy clustering. In Section~\ref{sec:discussion_conclusions}, we briefly place our results in context and provide our conclusions.

\section{The E-MOSAICS Simulations}
\label{sec:E-MOSAICS}

The E-MOSAICS (MOdelling Star cluster population Assembly In Cosmological Simulations within EAGLE) project combines the EAGLE galaxy formation model \citep{Schaye_2015,Crain_2015} with the sub-grid, semi-analytical MOSAICS model \citep{Kruijssen_2011, Pfeffer_2018}, which describes the formation, evolution and disruption of the entire star cluster population within the cosmic volume modelled. The simulations were run using the $N$-body TreePM smoothed particle hydrodynamics code \textsc{Gadget}~3, of which the most recent documentation can be found in \citet{Springel_2005}. The cosmological parameters assumed for the simulation discussed in this work are those determined by the \citet{PlanckCollab_2014}, most notably $\Omega_\mathrm{m} = 0.307$, $\Omega_\Lambda = 0.693$, $\Omega_\mathrm{b} = 0.04825$, $h = 0.6777$, and $\sigma_8 = 0.8288$.

The star formation prescription in EAGLE is calibrated to reproduce the empirical Kennicutt-Schmidt \citep{Schmidt_1959,Kennicutt_1998} relation \citep{Schaye_2008}. Furthermore, EAGLE includes feedback associated with star formation and black hole accretion that couples to the surrounding gas, using stochastic heating of the adjacent gas particles (see \citealt{Crain_2015,Schaye_2015} for details). The MOSAICS model employed in E-MOSAICS uses the physical model for the environmentally dependent cluster formation efficiency (CFE, \citealt{Bastian_2008}) from \citet{Kruijssen_2012} to determine the fraction of stellar mass assigned to one or multiple clusters during star formation. It also keeps track of the subsequent mass loss of these clusters, covering mass loss by stellar evolution, two-body relaxation-driven evaporation in a tidal field, and transient tidal perturbations (`tidal shocks'). Most of the cluster mass loss in the model is caused by tidal shocks \citep{Pfeffer_2018}. During post-processing, MOSAICS removes the clusters that were disrupted by dynamical friction. For further details on the model, see \citet{Kruijssen_2011, Pfeffer_2018}.

The E-MOSAICS simulations have proven to be successful at reproducing the properties of old and young star clusters in the local Universe \citep{Kruijssen_2019a, pfeffer19b, hughes20}. Among these, the simulated cluster populations reproduce the overabundence of massive red clusters, the `blue tilt' \citep{usher18}, the high-mass end of the GC mass distribution function \citep{Pfeffer_2018, Kruijssen_2019a}, and the distribution of metallicity across galaxy stellar mass \citep{Pfeffer+2023}. Furthermore, the simulations produce radial number density profiles of GCs which are also in very good agreement with observations for a wide range of galaxy stellar masses \citep{Reina-Campos_2021}. Besides, they have produced a fraction of stars contained in GCs \citep{Bastian_2020} and age-metallicity relations of GC systems \citep{Kruijssen_2019b,kruijssen20c, Horta_2021}, both of which are consistent with observed relations.

This work is one of the first to analyse a cosmologically representative volume evolved with the E-MOSAICS model \citep{Bastian_2020}. The simulation, which is given the index L034N1034 following the EAGLE nomenclature, evolves a periodic cube with a side length of $34.37\,\mathrm{cMpc}$ from a redshift of $z=127$ to $z=0$. The simulation initial conditions comprise $1034^3$ dark matter particles and an equal number of gas particles. The mass of each dark matter particle is $1.21 \times 10^6 \MSun$, and the initial mass of a gas particle is $2.26 \times 10^5 \MSun$. Star particles have an average mass of $2 \times 10^5 \MSun$. The model adopts the `Recalibrated' EAGLE model (for more details, see \citealt{Schaye_2015}) and has a mass resolution identical to the L025N0752 volume.

Over the course of the simulation, 29 snapshots between redshifts $z=20$ and $z=0$ were saved, but this work only uses the $z=0$ snapshot. Dark matter haloes and the galaxies within them are identified using a friends-of-friends algorithm \citep{Davis_1985} and subsequently the \textsc{Subfind} algorithm \citep{Springel_2001, Dolag_2009}, as described in \citet{Schaye_2015}. Galaxies located at the centre of potential of their respective dark matter haloes are identified as central galaxies, whereas all other galaxies belonging to the same dark matter halo are referred to as satellite galaxies. In total, the volume contains nearly 3000 galaxies with stellar masses $>2 \times 10^7 \MSun$. We exclude galaxies below $10^8 \MSun$ for the most part of this work. Below that mass, galaxies are composed of approximately 500 star particles or less, which makes them particularly susceptible to stochastic variations in the star formation rate (SFR) \citep[see e.g.][]{Davies_2021,Borrow_2023}, as well as to other shortcomings of the adopted star formation model.

E-MOSAICS includes several parallel realisations of the cluster population, obtained using various different cluster formation models. These models differ in how the CFE and the initial cluster mass function (ICMF) depend on the local environment, and have been explored by \citet{Pfeffer_2018,Reina-Campos_2019}. The `fiducial' cluster formation model, in which both quantities are described using physically-motivated, environmentally-dependent models \citep{Kruijssen_2012, Reina-Campos_2017}, is found to best reproduce observations of active star-forming galaxies in the Local Universe. Therefore, we adopt this fiducial model for our study. To save memory, clusters with an initial mass below $5 \times 10^3 \MSun$ are immediately discarded. Furthermore, star clusters are considered to be fully disrupted once their mass falls below $10^2\MSun$. 

To define globular clusters (GCs), we adopt a minimum mass threshold which depends on the stellar mass of the galaxy where the cluster is formed (see Table~\ref{tab:gc_masslims}). This follows \citet{Hughes_2022}, who find that the upper truncation mass $M_{\mathrm{c},*}$ of the GC initial mass function depends on the galaxy mass.
The lower limits are necessary to ensure that the fitted truncation mass in the Schechter mass function~\citep{Schechter1976} can be determined and does not exceed the actual mass of the most massive cluster in a galaxy \citep[see appendix A in][]{Hughes_2022}. Additionally, not only the GCs, but all the clusters in the volume are selected to have a metallicity [Fe/H] within the range specified in Table \ref{tab:metallicity_range}, where again the upper boundaries depend on the host galaxy's stellar mass $M_\star$. The exact values for these metallicity limits are taken from \citet{Reina-Campos_2021}, as are the metallicity cuts in Table \ref{tab:metallicity_cuts}, i.e.\ the threshold we define to distinguish between metal-rich and metal-poor clusters. The purpose of the galaxy mass-dependent upper metallicity limit for clusters is to remove artificially underdisrupted clusters. In E-MOSAICS, the lack of a cold interstellar medium (ISM) causes an underdisruption of clusters, as the destructive effect of tidal shocks from the cold ISM is underestimated. As a result, the predominantly metal-rich clusters that stay in the gas-rich disc for a longer duration are less likely to be disrupted compared to the real Universe (see appendix~D in \citealt{Kruijssen_2019a} for a detailed analysis).

\begin{table}
    \centering
    \caption{Lower mass limits used for the definition of GCs}
    \begin{tabular}{c c}
        \hline
        Galaxy Stellar Mass & Minimum GC Mass \\
        $\log(M_{\star,\,\mathrm{gal}}\,/\MSun)$ & $\MSun$\\
        \hline
        $< 9.0$ &  $10^4$ \\
        $\in [9.0, 10.0)$ & $3 \times 10^4$\\
        $\geq 10.0$ & $10^5$\\ \hline
    \end{tabular}
    \label{tab:gc_masslims}
\end{table}

\begin{table}
    \centering
    \caption{Metallicity range applied per galaxy stellar mass}
    \begin{tabular}{c c}
        \hline
        Galaxy Stellar Mass & Metallicity Range [Fe/H]\\
        $\log(M_{\star,\,\mathrm{gal}}\,/\MSun)$ & dex\\
        \hline
        $< 8.5$ &  $(-2.5,\, -1.0)$ \\
        $\in [8.5, 9.0)$ & $(-2.5,\, -1.1)$\\
        $\in [9.0, 9.5)$ & $(-2.5,\, -0.8)$\\
        $\in [9.5, 10.0)$ & $(-2.5,\, -0.5)$\\
        $\in [10.0, 10.5)$ & $(-2.5,\, -0.5)$\\
        $\in [10.5, 11.0)$ & $(-2.5,\, -0.5)$\\
        $\geq 11.0$ & $(-2.5,\, -0.3)$\\ \hline
    \end{tabular}
    \label{tab:metallicity_range}
\end{table}

In this work, we focus on the formation histories of stars, clusters, and globular clusters (GCs) as a function of the host galaxy's stellar mass. We define GCs as surviving clusters with a current mass larger than the minimum mass listed in Table~\ref{tab:gc_masslims}, while the general cluster population encompasses all surviving clusters. We analyse the galaxies across the full cubic volume with 34.37~comoving Mpc (cMpc) side length in the E-MOSAICS simulation for redshifts $0 \leq z \leq 10$.

\section{Formation histories of stars and stellar clusters in E-MOSAICS}
\label{sec:results}

We first present the median formation rates of stars, clusters, and GCs in central galaxies as a function of galaxy stellar mass. We opt specifically for the median formation history as the metric of interest, because it is insensitive to outliers, which here would mostly be galaxies where the environment has strongly affected their respective evolution. Such an environmental dependence would obscure the dependence on galaxy mass that we focus on here. When showing the GC formation history, we include all clusters above the minimum mass limit listed in Table~\ref{tab:gc_masslims}.
This is consistent with the premise of the models used in this work, which is that GCs are the result of intense star and cluster formation throughout cosmic history \citep[e.g.][]{kruijssen15b, Pfeffer_2018}. When we later in the paper specifically consider the formation history of `proto'-GCs, meaning clusters with initial masses above the threshold given by Table~\ref{tab:gc_masslims} , this will be indicated as such (see Section~\ref{subsec:formrate_densities}).

\subsection{Dependence of formation histories on galaxy mass}
\label{subsec:formhists_per_galmass}

\begin{figure*}
    \centering
    \includegraphics[scale=0.8]{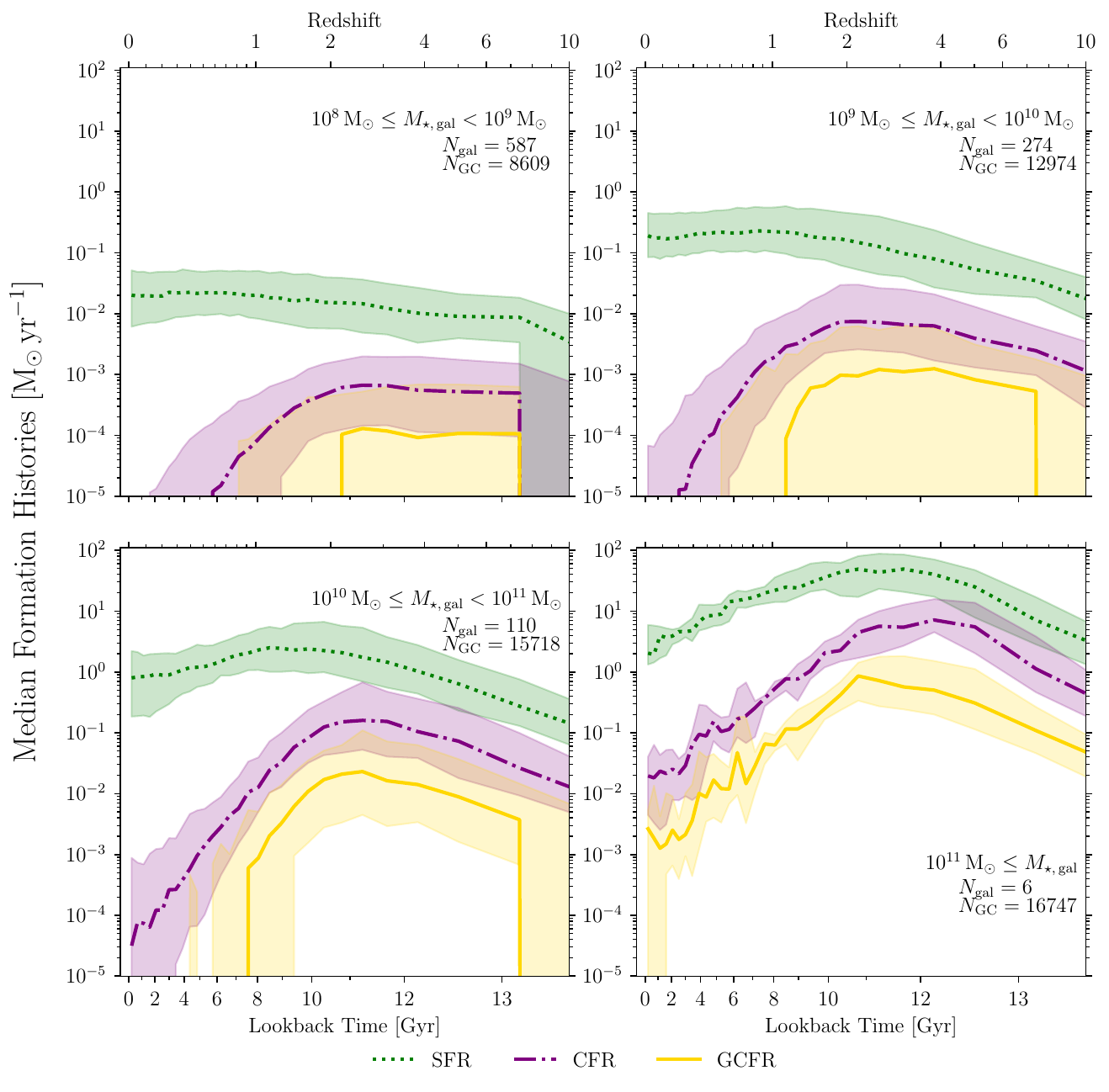}
    \caption{Median formation histories for stars (\textit{green}), clusters (\textit{purple}) and globular clusters (\textit{yellow}) sorted by galaxy stellar mass as indicated in each panel. Shaded areas in the respective colours indicate values between the 16th and 84th percentiles. The total numbers of galaxies and (proto-)GCs are annotated in each respective panel. The peak formation rates differ by $3{-}4$ orders of magnitude between the least and the most massive galaxies. The peak of GC formation typically precedes the peak of star formation, with the time interval increasing towards lower galaxy masses, from no discernible offset at $\mgal>10^{11}~\MSun$ to $\Delta t=6~\Gyr$ at $\mgal<10^9~\MSun$. This offset is mostly driven by a later peak in the SFR, whereas the peak in the GC formation history is nearly constant at $z=2{-}4$.}
    \label{fig:formhists}
\end{figure*}

We separate the population of simulated central galaxies into four stellar mass bins. \autoref{fig:formhists} shows the formation rates of stars, clusters, and GCs as a function of both lookback time and redshift $z$ for each of these bins. The six most massive galaxies contain about 30~per~cent of all GCs, with a similar amount located in the $\sim100$ galaxies with Milky Way-like masses ($M_{\star,\mathrm{gal}} = 10^{10}-10^{11}\MSun$).

All of the formation histories shown in \autoref{fig:formhists} show the characteristic peaked shape known from the cosmic star formation history \citep[e.g.][]{Madau_2014}. However, the redshift (or lookback time) at which the peak is reached depends on the galaxy mass bin and tracer. The SFR shows the strongest galaxy mass dependence of the peak redshift, with the peak SFR being reached at $z_{\rm peak,SFR}=\{0.5, 1, 1.5, 2.5\}$ for the four galaxy mass bins with lower mass galaxies reaching their peak later than massive ones, but the weakest decline from the peak SFR to the SFR at $z=0$. For galaxies with $\mgal<10^{10}\MSun$, the post-peak decline is almost unnoticeable.

Both the cluster formation rate (CFR) and the globular cluster formation rate (GCFR) exhibit a characteristic peaked shape. The formation rates increase from early times until they reach a peak around redshift $z=2{-}4$ and subsequently decrease until the present time. The peak redshift depends on the galaxy mass. For the CFR, we find that $z_{\rm peak,CFR}=\{2.5, 2, 2.5, 4\}$, whereas for the GCFR in order of increasing galaxy mass  $z_{\rm peak,GCFR}=\{4, 3.5, 2.5, 2\}$.\footnote{An interesting feature of the highest-mass galaxies is that in contrast to the other galaxy mass bins, the peak of the CFR ($z\approx4$) precedes the peak of the GCFR ($z\approx2$). We further investigate this shift between the two peaks in Section~\ref{subsec:formrate_densities}.} In the lowest mass bin, the GCFR nominally peaks at $z\approx 4$, but the formation history is nearly constant between $z=2{-}7$. The increase of the peak CFR with galaxy mass coincides with observations of UV clumps, of which the incidence also peaks earlier in galaxies with higher mass~\cite{Guo+2015,Shibuya+2016} and which may provide a favourable environment for cluster formation. An interesting observation is that the peak formation redshift increases with galaxy mass for stars, but decreases with galaxy mass for GCs. As a result, GCs form during a markedly different epoch of cosmic history than most of the stars in low-mass galaxies, but simultaneously with most of the stars in the most massive galaxies. Because GC formation requires high gas pressures \citep{elmegreen97,kruijssen15b,keller20}, this suggests that low-mass galaxies cannot sustain the pressures required for GC formation at $z\lesssim 2$, while they do continue to form stars. Massive galaxies sustain the required pressures long enough that GC formation ends up following most of the star formation.

The pressure requirements for efficient cluster formation \citep{Kruijssen_2012} also impact the shut-off of cluster formation within the galaxies in our simulation. GC formation has ceased in galaxies with masses $\mgal<10^{11}\MSun$, and in galaxies with $\mgal<10^{10}\MSun$ cluster formation has mostly ceased in general. This implies that the vast majority of stars that form in these galaxies are field stars \citep[as is found in observations, see e.g.][]{adamo20}. Having said that, we reiterate that our predictions regarding young massive cluster formation are limited by the upper metallicity limit we adopt for clusters (see Table~\ref{tab:metallicity_range}) to limit the effects of underdisruption present in this metallicity range. Although \citet{Pfeffer+2023} show that E-MOSAICS is able to largely reproduce observed cluster abundances even for higher metallicities, it is impossible within the scope of this work to distinguish between metal-rich clusters that should form and clusters that only remain due to underdisruption. We refer to Appendix~\ref{appendix} for a quantitative discussion of how the metallicity cuts impact the reported formation histories.

\subsection{The epoch of high star and (globular) cluster formation activity}
\label{subsec:period_high_activity}

\begin{figure*}
    \centering
    \includegraphics[scale=0.8]{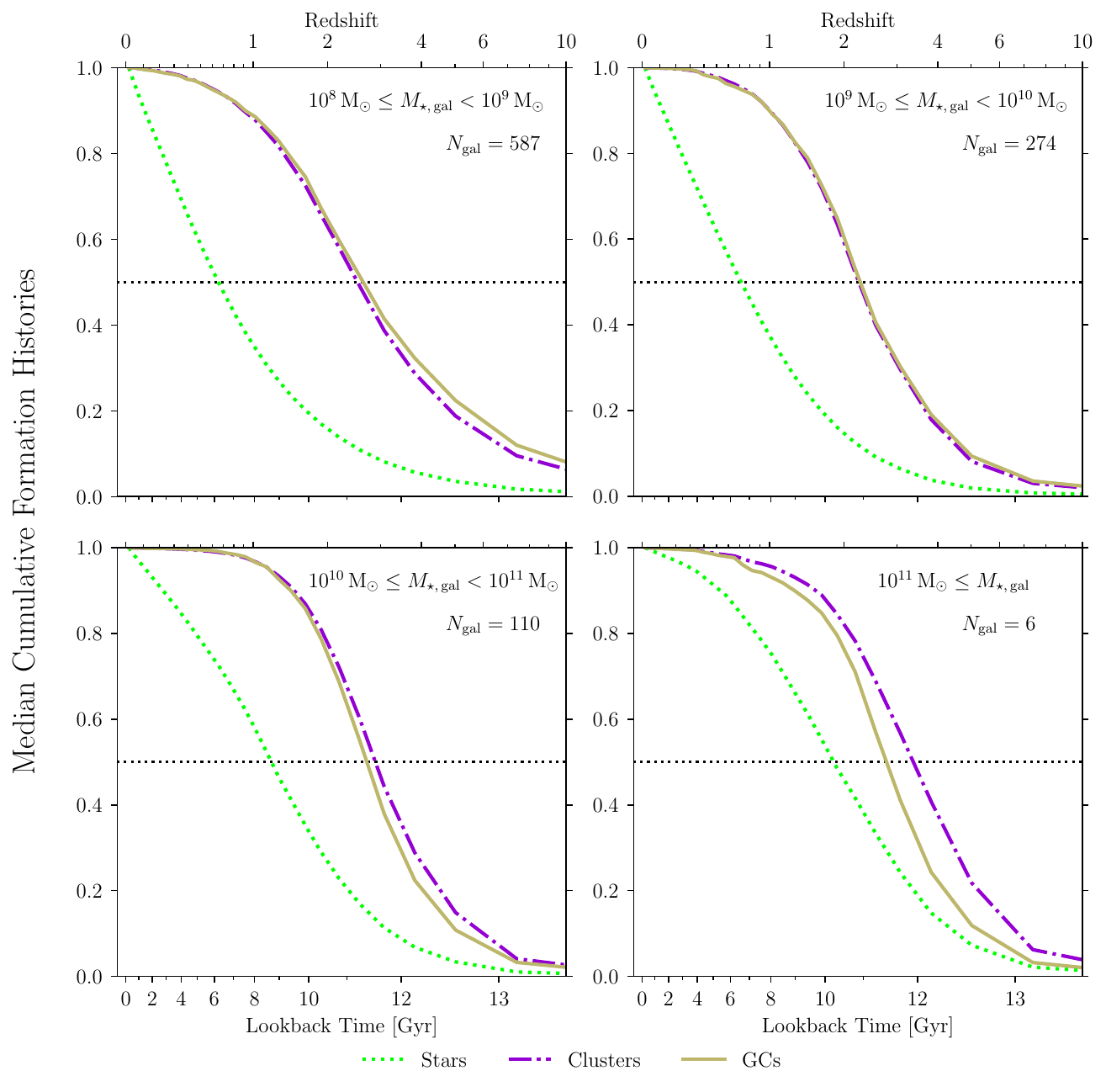}
    \caption{Normalised cumulative formation histories of stars (\textit{green}), clusters (\textit{violet}) and GCs (\textit{khaki}) in central galaxies, for different galaxy mass bins (different panels). Each panel shows how much of the population has formed by the respective lookback time, in units of the total mass at $z=0$. The black dotted line is placed at a percentage of 50 per~cent. Periods of high formation activity are indicated by a steep gradient.}
    \label{fig:cumulative_formhists}
\end{figure*}

All galaxies with present-day mass above $10^9 \MSun$ show a peak epoch of star formation activity, and the peak becomes less defined in lower-mass galaxies. These periods of relative high formation activity need not exactly coincide, but generally they culminate around $z\sim 2$. To pinpoint this epoch more precisely, Fig.~\ref{fig:cumulative_formhists} shows the cumulative formation histories of stars, clusters, and GCs as a function of both lookback time and redshift. Here, `cumulative' means that we divide the mass formed by a given time by the total mass at $z=0$, such that periods with an enhanced formation rate manifest themselves as a strong gradient. For reference, we add a horizontal dotted line at 50 per~cent, i.e.\ half of the total mass of each population.

From Fig.~\ref{fig:cumulative_formhists}, we learn that galaxies with $M_{\star, \mathrm{gal}} < 10^{10}\MSun$ have the youngest stellar populations, such that about 60 per~cent of all their stars formed at $z\leq1$, consistent with Fig.~\ref{fig:formhists}. The cumulative SFR at the lowest galaxy masses reaches its steepest slope at $z=0$, reflecting the observation from Fig.~\ref{fig:formhists} that the SFR is currently highest. By contrast, the cumulative formation rates of clusters and GCs reach their strongest gradient for $z=2-3$. Interestingly, the simulation predicts that galaxies with stellar mass $10^8 \MSun \leq M_{\star, \mathrm{gal}} < 10^9\MSun$ host a non-negligible fraction of clusters that formed in the very early universe $z>10$, constituting up to 10 per~cent of the total surviving clusters' mass. This fraction is negligible at higher galaxy masses owing to more efficient disruption.

Fig.~\ref{fig:cumulative_formhists} also highlights the differences between the different mass bins. Towards higher galaxy masses, the redshift evolution of the CFR and GCFR closely follow the shape of the SFR, with the peak SFR occurring at $z=1{-}2$ for massive galaxies with $M_{\star, \mathrm{gal}}>10^{10} \MSun$, and the majority of stars in these galaxies forming at $z\geq 1$. Likewise, the slope of the cumulative CFR and GCFR in these high-mass galaxies increases slowly at first, until a period of peak formation activity at around $z=3$, after which the formation rate decreases again.

Interestingly, the median formation redshift (the intersection with the horizontal dotted line) approximately coincides with the steepest gradient for many cumulative formation rates. This means that galaxies formed stars, clusters, and GCs most intensively when about half of each respective population was in place.

Quantitatively, half of the stars in low-mass galaxies ($<10^{10}\MSun$) were in place at $z\simeq0.7$. In more massive galaxies, this point was reached at $z=1{-}2$. By contrast, half of the star cluster populations were in place at $z=\{2.5, 2.2, 2.9, 3.3\}$ for the four galaxy mass bins, in order of increasing galaxy mass. Notably, the median formation redshift of GCs also increases with galaxy mass, but does so much more weakly than that of all clusters -- half of the GCs were in place at $z=2.3{-}2.8$ across all galaxy mass bins. This means that the entire cluster population generally becomes older the more massive the host galaxy is, whereas GCs follow a much weaker trend between their age and the host galaxy mass.

\subsection{Formation Histories of Metal-Poor and Metal-Rich Globular Clusters}
\label{subsec:gc_formhist_metallicity}

\begin{table}
    \centering
    \caption{Cuts used for defining `metal-poor' and `metal-rich' GCs}
    \begin{tabular}{c c}
        \hline
        Galaxy Mass & Metallicity cut [Fe/H]\\
        $\log(M_{\star,\,\mathrm{gal}}\,/\MSun)$ & dex\\
        \hline
        $< 8.5$ &  $-1.2$ \\
        $\in [8.5, 9.0)$ & $-1.2$\\
        $\in [9.0, 9.5)$ & $-1.2$\\
        $\in [9.5, 10.0)$ & $-1.1$\\
        $\in [10.0, 10.5)$ & $-1.0$\\
        $\in [10.5, 11.0]$ & $-0.9$\\
        $\geq 11.0$ & $-0.8$\\
        \hline
    \end{tabular}
    \label{tab:metallicity_cuts}
\end{table}

It is thought that the properties of GCs are shaped by the galactic environment at the time of their formation. As a result, their lifecycles may depend strongly on their natal galaxy mass, which at the present day is best traced by the GC metallicity \citep[e.g.][]{kruijssen15b}. We express the metallicity as [Fe/H], and consider the median GC formation histories in different GC metallicity bins for each galaxy mass interval.

We follow common practice to divide the GC population into `metal-poor' and `metal-rich' GCs \citep[e.g.][]{brodie06}. Moreover, we follow \citet{Reina-Campos_2021} in describing the metallicity threshold between metal-poor and metal-rich GCs within E-MOSAICS as a function of the host galaxy's stellar mass (also see \citealt{peng06}), as listed in Table~\ref{tab:metallicity_cuts}. The reason for treating the metallicity cut as a variable is the same as to why we adopt a variable metallicity range for our clusters, i.e.\ the typical GC metallicity increases with galaxy mass (see Section~\ref{sec:E-MOSAICS} as well as Table~\ref{tab:metallicity_range}).

\begin{figure*}
    \centering
    \includegraphics[scale=0.8]{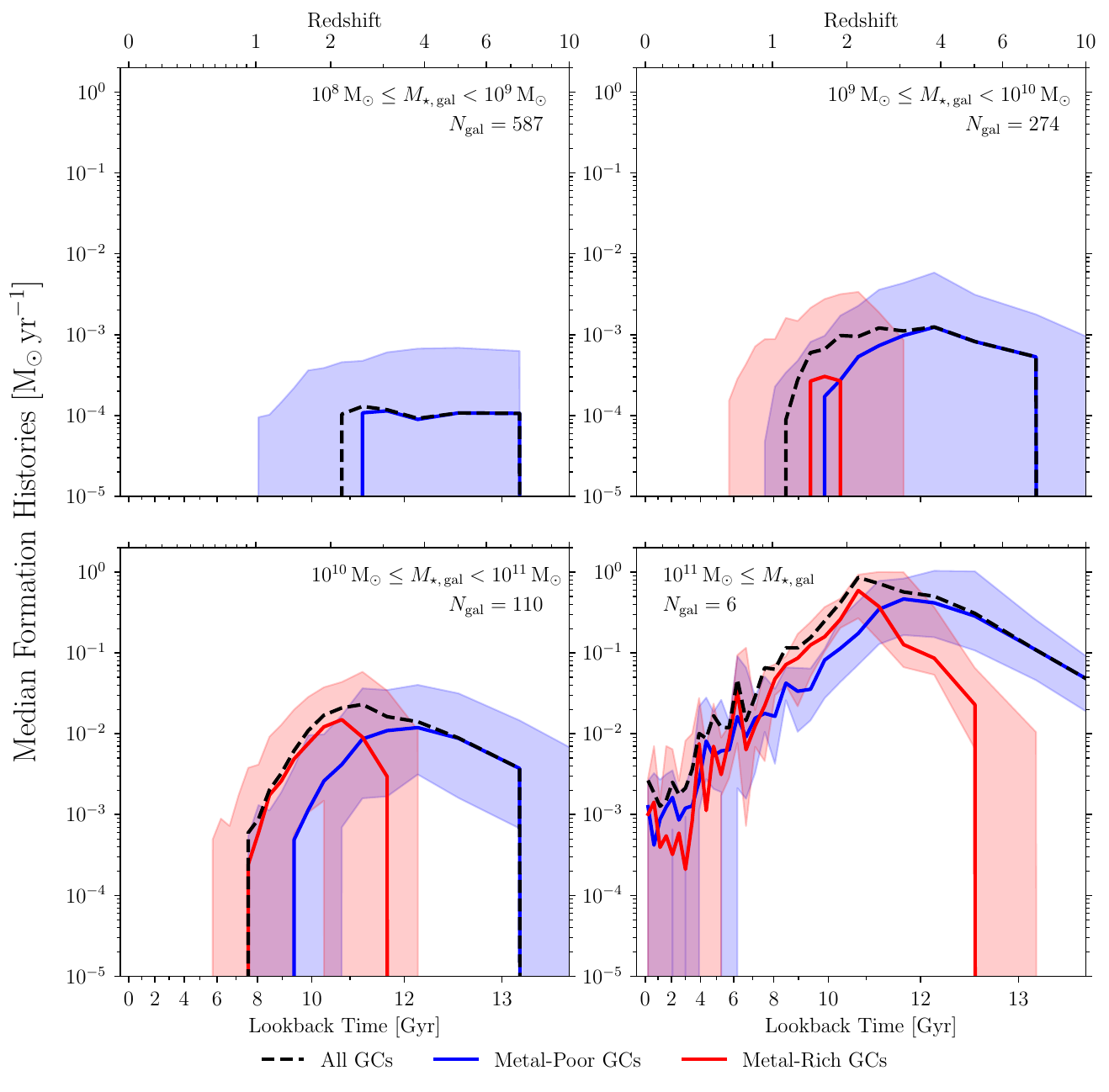}
    \caption{Median formation histories of metal-poor (\textit{blue}) and metal-rich (\textit{red}) GCs, as defined in Table~\ref{tab:metallicity_cuts}. The shaded areas correspond to the range between the respective 16th and 84th percentiles as in Fig.~\ref{fig:formhists}. For reference, we include the median formation rate for all GCs (\textit{black}) from Fig.~\ref{fig:formhists}. This figure illustrates the chemical enrichment of the host galaxies during GC formation, as GC formation shifts from metal-poor to predominantly metal-rich clusters.}
    \label{fig:gc_metallicity_formhists}
\end{figure*}

The metallicity-separated GC formation histories are shown in Fig.~\ref{fig:gc_metallicity_formhists} for each of the usual four galaxy mass bins. For reference, we also include the total median GC formation histories from Fig.~\ref{fig:formhists} as dashed black lines. The figure shows the delayed formation of metal-rich GCs compared to metal-poor GCs across all galaxy mass bins, which reflects the ongoing chemical enrichment of the host galaxy. While metal-rich GCs are formed for all galaxies with masses $>10^9~\MSun$, the period of metal-rich GC formation begins earlier in higher mass bins, increasing from $z\approx3$ at the low end of the galaxy mass range to $z \approx 7$ at the very highest masses. This reflects the well-known observation that chemical enrichment proceeds faster in more massive galaxies \citep{CidFernandes_2007}.

The formation of metal-poor GCs ceases around $z=1$ for all galaxies with masses $<10^{11}~\MSun$. At the highest galaxy masses (bottom-right panel), the formation of metal-poor clusters continues until $z=0$, at a rate similar to that of metal-rich GCs. This reflects the accretion of small, metal-poor galaxies and is discussed further in Section~\ref{sec:discussion_conclusions}.

As a note of caution, we remind the reader that the GC samples from E-MOSAICS adopt a metallicity cut to exclude the highest-metallicity GCs and remedy the underdisruption of GCs. As a result, our ability to make definitive predictions for the formation of the most metal-rich GCs is limited, especially toward lower redshifts when GCs with the highest metallicities are formed. We refer to Appendix~\ref{appendix} for a quantitative discussion of how the metallicity cuts impact the reported formation histories.

\subsection{Dependence on Galaxy Mass}
\label{subsec:galaxy_mass_dependence}

\begin{figure}
    \centering
    \includegraphics[width=\columnwidth]{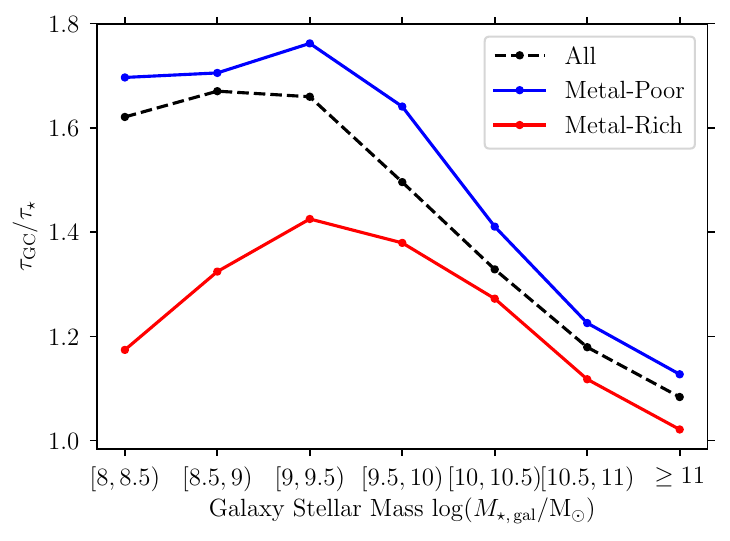}
    \caption{Median age of GCs divided by the median age of all stars in central galaxies, as a function of galaxy mass. Red and blue lines indicate the same quantity for different metallicity cuts (see Table~\ref{tab:metallicity_cuts}). For galaxies with masses above $10^{9.5}\MSun$, the median ages of stars approach the median ages of the GCs, while below that threshold only metal-rich GCs have ages similar to those of the stars. Metal-poor GCs formed well before most of the stars at low galaxy masses.}
    \label{fig:tgc_per_tstar}
\end{figure}

In order to further visualise the difference in the GC formation histories between different galaxy masses, and to quantify differences relative to all star formation in the host galaxy, Fig.~\ref{fig:tgc_per_tstar} shows the median age $\tau_\mathrm{GC}$ of the galaxies' GC population divided by the median age $\tau_\star$ of their stellar population. This quantity gives a relative measure of how much earlier the GCs formed relative to the stellar content of the host galaxy. We consider this relative formation lookback time as a function of galaxy mass, both for all GCs and separately for metal-rich and metal-poor GCs.

Taking a look at the general GC population first, we see that the median GC age is $1.6{-}1.7$ times older than the median stellar age in galaxies up to $10^{9.5}\MSun$. The ratio $\tau_\mathrm{GC}/\tau_\star$ decreases steadily for more massive galaxies, down to $\tau_\mathrm{GC}/\tau_\star\sim1.1$ in galaxies with $\log(M_{\star ,\mathrm{gal}} / \mathrm{M_{\odot}}) \geq 11$. As we discuss below, this trend is not due to GCs being generally younger in high-mass galaxies, but instead the stellar population is older on average with growing galaxy mass. This phenomenon, commonly referred to as downsizing \citep{Thomas_2005, Neistein_2006}, is well known from observations, where more massive galaxies also host older stellar populations. It has also been reproduced in previous numerical simulations, \citep[e.g.][]{Oser_2010}.

Metal-poor GCs follow the same general trend described above for all GCs, except that the line for metal-poor GCs is at most $0.05{-}0.15$ higher than that of the entire GC population. This shows that metal-poor GCs simply formed slightly before the bulk of the GC population by some fixed relative time offset. The metal-rich GCs exhibit an interesting difference from these trends at galaxy masses below $10^{9.5}\MSun$, where $\tau_\mathrm{GC}/\tau_\star$ drops from $\sim1.4$ at the high end of this mass range to $<1.2$ at the lowest masses. As before, this downturn is driven by the stars being older in these lowest-mass galaxies, which is not what we would expect from downsizing. As we discuss below, the stellar ages in our simulation follow the prediction of downsizing only for masses above $\sim10^9 \MSun$, where they hit a threshold below which downsizing is no longer evident. This might be related to the resolution of the simulations, as the stochasticity of the star formation histories increases considerably for the lowest-mass galaxies.

\begin{figure*}
    \centering
    \includegraphics[scale=0.74]{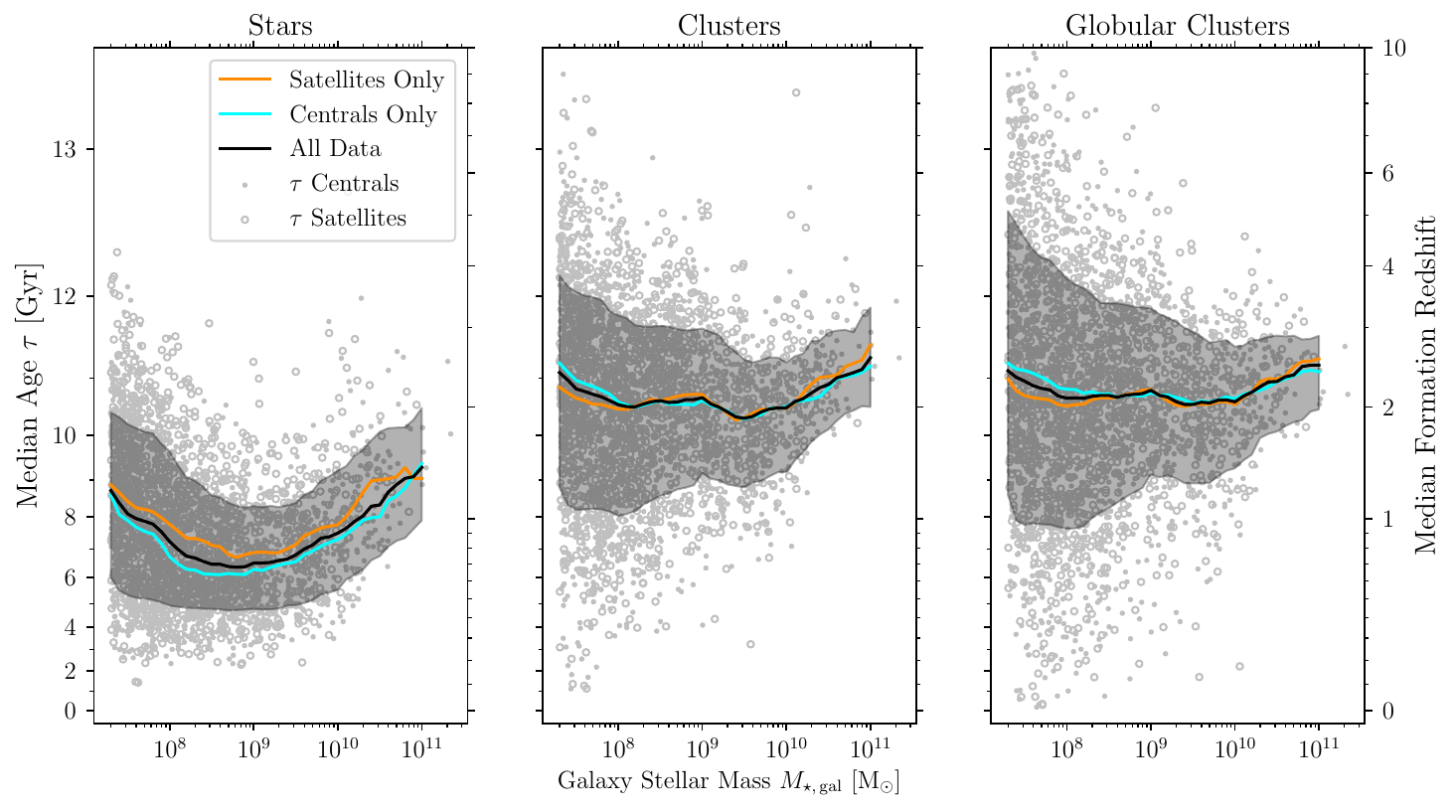}
    \caption{Median ages of stars (\textit{left}), clusters (\textit{middle}) and GCs (\textit{right}) as a function of galaxy stellar mass and given in both lookback time and redshift. Central galaxies are indicated as full grey dots, whereas satellites are grey circles. We show the running median with a bin width of 1~dex for all galaxies in the volume, as well as the 16th and 84th percentiles (black line and shaded area, respectively). Additionally, we show the running medians for centrals (\textit{blue}) or satellites (\textit{orange}) only. Stars in satellites formed before stars in centrals, which is related to slightly earlier quenching. This environmental dependence is not visible for the clusters and GCs, because these formed before the satellites were strongly affected by their current environment.}
    \label{fig:age_scatter}
\end{figure*}

We now consider how the formation histories depend on the large-scale galactic environment. Fig.~\ref{fig:age_scatter} displays in three separate scatter plots the median ages of stars, clusters, and GCs of all galaxies, both centrals and satellites, as functions of stellar mass. Galaxies without GCs are omitted from the right-hand panel.  These galaxies almost all have masses $M_\star < 10^{8}\MSun$, with the same incidence for centrals and satellites. We find that there is a large spread of ages over several Gyr for all three populations, especially towards lower galaxy masses. To better visualise the trends, we apply a running median with a width of 1~dex in mass and also show the range between the 16th and 84th percentiles. This procedure is then repeated for satellites and centrals independently.

Starting with the median ages of the stars, we see a minimum of the median age $\tau$ for galaxies with masses of $10^9\MSun$, where $\tau\approx 6.5\Gyr$, and an increase towards both lower and higher masses, up to $\tau\approx 9\Gyr$. The scatter increases slightly towards low-mass galaxies. Both satellites and centrals follow this trend, but the stellar populations in satellite galaxies are systematically older by $0.5-1\Gyr$. This is likely related to the environmental quenching of satellites \citep{Guo_2011,Wetzel_2013,Pasquali_2015}. The trend of increasing median age towards low galaxy masses is somewhat unexpected in the context of observations of dwarf galaxies in the Local Group \citep{Weisz_2014}, and it may be a result of EAGLE's limited resolution in low-mass galaxies \citep{Schaye_2015}. For galaxies with masses above $10^9\MSun$, Fig.~\ref{fig:age_scatter} shows that downsizing is indeed reproduced in the EAGLE model.

Unlike the stars, the clusters and GCs in Fig.~\ref{fig:age_scatter} do not exhibit any significant trends with galaxy mass or environment. As discussed above, the median ages of both populations always correspond to $z=2{-}3$. Only in galaxies with $M_\star \gtrsim 10^{10}\MSun$, the median ages of clusters and GCs increase from $z \approx 2.0$ to $z \approx 2.5$. Additionally, the median ages of cluster and GC populations in centrals and satellites are not systematically offset, contrary to those of the stars. The environmental dependence of the stellar ages is not visible for the clusters and GCs, because these formed before the satellites were strongly affected by their current environment. Finally, we see that the scatter of the median ages of clusters and GCs generally grows towards lower $M_\star$, contrary to the nearly constant scatter in stellar median ages. This suggests that cluster formation (and especially GC formation) is more stochastic than star formation, and the scatter of the median ages of clusters and GCs in low-mass galaxies traces this stochasticity. Such behaviour is unsurprising, because clusters (and especially GCs) represent mass quanta that are a factor of $10^3{-}10^5$ larger than stars. As a result, their formation is governed by small-number statistics, and increasingly so towards lower-mass galaxies.

\subsection{The Formation Rate Densities}
\label{subsec:formrate_densities}

\begin{figure*}
    \centering
    \includegraphics{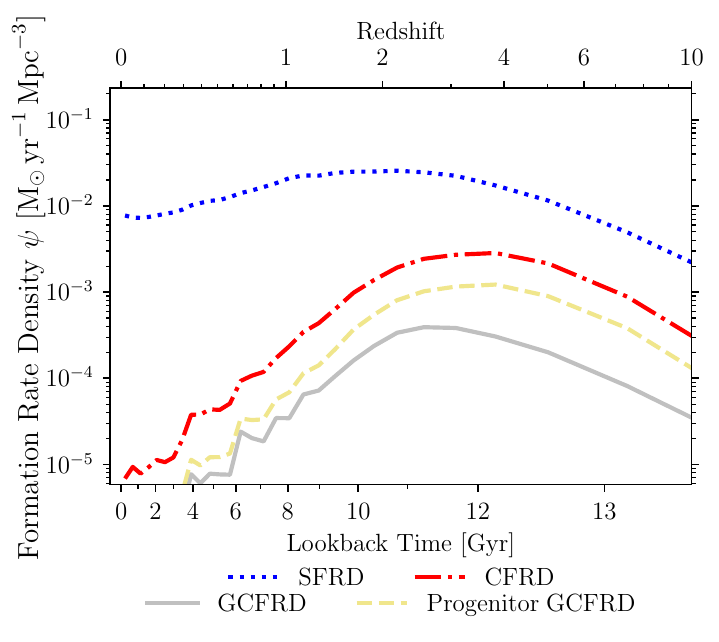}
    \caption{Total formation rate densities (i.e.\ per unit comoving volume) for the different populations: stars (\textit{blue}), clusters (\textit{red}), and GCs (\textit{grey}), as well as clusters with an initial mass larger than $10^5 \MSun$ (\textit{yellow}), which we refer to as `progenitor GCs'. Not all progenitors survive to be GCs, especially if they form early. As a result of this survivor bias, the GC formation rate density peaks after the cluster and progenitor GC formation rate densities.}
    \label{fig:formhist_densities}
\end{figure*}

In Fig.~\ref{fig:formhist_densities}, we show the comoving formation rate densities in the entire 34~Mpc box, obtained by dividing the formation rates by the comoving volume. As before, we provide the formation rate densities of stars (SFRD), clusters (CFRD), GCs (GCFRD) and progenitor GCs (or `proto-GCs') (progenitor GCFRD, i.e.\ clusters with an initial mass $\geq 10^5 \MSun$, which may or may not evolve into GCs eventually). We observe here immediately that the integrated formation histories follow a similar qualitative behaviour as seen for the individual mass bins in Fig.~\ref{fig:formhists}, i.e.\ the formation rate densities peak at some redshift in the range $z=2{-}4$. The individual peak redshifts best match those seen at the highest galaxy masses ($M_\star \geq 10^{11}\MSun$), as the SFRD peaks at $z\approx2$, the GCFRD peaks at $z\approx2.5$, and the CFRD and progenitor GCFRD peak at $z\approx4$ \citep[cf.\ also][]{Guo+2015,Shibuya+2016}. The SFRD exhibits a strong qualitative difference relative to the formation rate densities of clusters and (proto-)GCs, as it does not exhibit the factor-of-100 drop from $z\approx3$ to the present day. This reflects the changing cosmic conditions, in which stars continue to form, but the high gas pressures that are conducive to (massive) cluster formation are favoured at high redshift.

The CFRD and progenitor GCFRD show a constant offset of approximately 0.5~dex (or a factor of 3). This is to be expected for a power-law initial cluster mass function with a slope of $-2$ between $10^2{-}10^6~\MSun$. As a result, the CFRD and progenitor GCFRD peak at a similar redshift. The fact that the GCFRD peaks at a lower redshift reflects a survival bias -- in order to be classified as a GC at $z=0$, the cluster needs to have a mass above $10^5~\MSun$, which favours clusters that formed later as these will have lost less mass through stellar evolution, tidal shocks, and dynamical friction. This is an important observational prediction: by comparing the progenitor GCFRD (observed with e.g.\ \emph{JWST} and the upcoming generation of thirty-metre class telescopes) to the age distributions of surviving GCs at $z=0$ (e.g.\ with upcoming wide-field spectroscopic surveys), this $1{-}2$~Gyr offset in the median may be observable.

We deliberately focus on differences between the various formation rate densities. The EAGLE galaxy formation model adopted here produces a cosmic SFRD that is a factor of a few below the observed SFRD from \citet{Madau_2014}. The reason behind this offset is that the galaxy stellar mass function of EAGLE is lower for galaxies with stellar masses $10^{10}\MSun \leq M_\star \leq 10^{11}\MSun$ and slightly higher above $10^{11}\MSun$ compared to observations \citep{Schaye_2015,Furlong+2015}. Due to the dominant contribution of $L^{\star}$ galaxies to the mass function, the simulated volume underpredicts the observed total stellar mass in the real Universe. This may also contribute to the similarity between the volume-integrated formation rate densities in Fig.~\ref{fig:formhist_densities} and those of the highest-mass galaxies in Fig.~\ref{fig:formhists}, as Milky Way-mass galaxies are under-represented in the simulations. As Fig.~\ref{fig:formhists} indicates, a larger proportion of Milky Way-mass galaxies would shift the peaks of the formation histories to slightly lower redshifts. Secondly, the volume size contributes to a `flatter' shape of the SFRD compared to \citet{Madau_2014}, since the volume is too small to pick up early star-forming peaks. This becomes particularly evident in \citet{Furlong+2015}, where the SFRD is shown for two volumes in EAGLE, for 25~Mpc and for 100~Mpc side lengths in figs.~B1 and~4, respectively.  However, the differences between the various formation histories that we report here would still persist if these two points were addressed.

\section{Conclusions}
\label{sec:discussion_conclusions}

This work provides insight into the formation histories of stars, clusters, and GCs over a representative cosmological volume of $34^3$~cMpc$^3$ in the E-MOSAICS simulations \citep{Pfeffer_2018,Kruijssen_2019a}. As a reminder, we use `GCs' to refer to clusters that survive to $z=0$ with a final mass of at least $10^5~\MSun$. The main results of this work are as follows.
\begin{enumerate}
    \item We find that the formation history of GCs depends considerably on the mass of their host galaxy. The more massive a galaxy is, the older its cluster population is predicted to be. The peak of GC formation typically precedes the peak of star formation, with the time interval increasing towards lower galaxy masses, from no discernible offset at $\mgal>10^{11}~\MSun$ to $\Delta t=6~\Gyr$ at $\mgal<10^9~\MSun$. This offset is mostly driven by a later peak in the SFR, whereas the peak in the GC formation history is nearly constant at $z=2{-}4$. At recent times ($z<1$), GC formation has nearly ceased compared to the high formation rates in the early Universe. GCs began to form at very high redshift, with GC formation at $z>10$ constituting up to 10~per~cent of all the GCs formed in galaxies with present-day stellar masses of $\mgal<10^{9}~\MSun$. This provides important context for upcoming \emph{JWST} observations of the proto-GC population at extremely high redshift.
    \item As expected, the peak of metal-poor GC formation precedes the peak of metal-rich GC formation (defined according to the cuts in galaxy mass-GC metallicity space from Table~\ref{tab:metallicity_cuts}). The age difference between the two populations increases towards lower galaxy masses. In the most massive galaxies ($\mgal>10^{11}~\MSun$) the difference is small, with metal-poor GC formation peaking at $z\sim3$ and metal-rich GC formation peaking at $z\sim2$, after which both metal-rich and metal-poor GCs form at similar rates until the present day.
    \item The median age of GCs ($\tau_{\rm GC}$) is 10--60~per~cent larger than the median age of stars in a galaxy ($\tau_{\star}$). This number decreases with galaxy mass, with $\tau_{\rm GC}/\tau_{\star}=1.6$ for galaxies with $\mgal<10^{9.5}~\MSun$, and $\tau_{\rm GC}/\tau_{\star}=1.1$ for the most massive galaxies ($\mgal>10^{11}~\MSun$).
    \item We consider the formation histories of GCs in the context of galaxy evolution. We look for any apparent `downsizing', i.e.\ the shift of formation activity from high-mass to low-mass galaxies towards lower redshifts. Due to their old ages, GCs do not exhibit a significant downsizing effect, and thus their median age does not depend very strongly on the host galaxy mass. There is also no discernible difference between the ages of GCs in central and satellite galaxies. Stars in satellites formed before stars in centrals, which is related to satellite quenching. The reason why GCs do not exhibit this offset is that they formed before the satellites were strongly affected by their current environment.
    \item The GC formation rate density across the entire $34^3$ cMpc$^3$ volume peaks at $z\approx2.5$, shortly before the peak star formation rate density ($z\approx2$), but well after the peak of the cluster formation rate density ($z\approx4$) and the peak {\it progenitor} GC formation rate density ($z\approx4$; where progenitor is defined as any cluster with a mass larger than $10^5~\MSun$ at the time of its formation). The offset between the peak of GC formation and GC progenitor formation is the result of a survivor bias, where GCs that survive to $z=0$ are more likely to have formed recently. After the peak of GC formation, the GC formation rate drops sharply due to changing cosmic conditions, which become unfavourable to GC formation at lower redshifts due to decreasing gas pressures.
\end{enumerate}

In addition to E-MOSAICS, there are many simulations that attempt to describe globular cluster formation in the context of galaxy formation \citep[e.g.][]{Mistani+2016,ChoksiGnedin_2019,Halbesma+2020,Doppel+2021,Reina-Campos_2022,Grudic+2023}, but there are few that sample the required large-number statistics within a cosmologically representative volume over the full course of a Hubble time. Using a semi-analytic model for the formation of GC populations, \citet{ChoksiGnedin_2019} also studied the cosmic GC formation history (predicting similar quantities as in our Section~\ref{subsec:formrate_densities} and Fig.~\ref{fig:formhist_densities}). Their model, as set forth in \citet{Choksi_2018}, uses data from the dark-matter-only, \emph{Illustris-1-dark} simulation \citep{Vogelsberger_2014, Nelson_2015}, which employs different physical models to EAGLE. In fig.~5 of \citet{ChoksiGnedin_2019}, the authors find that the GC formation rate density in their model peaks significantly earlier than the star formation rate density from \citet{Madau_2014}, i.e.\ at $z=4-6$. This is also considerably earlier than our prediction for the peak of cosmic GC formation, which we attribute to the direct treatment of baryons in E-MOSAICS, whereas these are treated semi-analytically in the \citet{ChoksiGnedin_2019} model. Finally, they also distinguish the (progenitor globular) cluster formation rate from the surviving GC formation rate. Our results are qualitatively in agreement, and their surviving GC formation history also peaks more recently compared to the progenitor GC formation history. While quantitative differences may exist, it looks like the relative shifts in peak formation rates between different components of galaxies are firm predictions of these models, independently of the adopted methodology.

Similarly, we are able to reproduce the results for the fiducial GC formation history of \citet{Reina-Campos_2019}, who model the formation rates of stars, clusters, and GCs in 25 Milky Way-like galaxies from the earlier suite of E-MOSAICS zoom-in simulations \citep{Pfeffer_2018,Kruijssen_2019a}. Their fiducial model matches the model that we adopt here. In particular, we reproduce the shape of the GC formation history for Milky Way-like galaxies (see the lower left panel of Fig.~\ref{fig:formhists}), with a peak between redshifts $z=2$ and~$3$, but using 110 galaxies rather than 10. Hence, the result that the formation of surviving GCs in Milky Way mass galaxies approximately coincides with active star formation during the same period holds up with vastly increased statistics.

Perhaps most importantly, the results of this work (and others, e.g.\ \citealt{kruijssen15b,ChoksiGnedin_2019,keller20,Horta_2021}) form an important counterpoint to models in the literature that assume special conditions for GC formation at redshifts $z>6$ \citep[e.g.][]{trenti15,boylankolchin17,madau20}. The E-MOSAICS series of papers has shown that no special conditions for GC formation at very high redshift are needed to reproduce the observed demographics of massive clusters across cosmic time, but instead GCs can be seen as natural by-products of intense star formation throughout cosmic history. \autoref{fig:formhist_densities} shows that the GCs in E-MOSAICS formed at any time that the environmental conditions enabled intense star formation and the subsequent survival of the resulting clusters. This means that the GC formation rate at $z=10$ already greatly exceeded the GC formation rate at the present day. With ongoing observations coming from \emph{JWST}, we will witness an increased volume of (proto-)GCs observed in high-redshift galaxies. As these observations accumulate to enable statistical studies, the results of this paper provide a contextual framework for helping to interpret the observations, and eventually may represent testable predictions for the history of GC formation across the galaxy population.

\acknowledgments
\noindent\textit{Acknowledgments:} J.M.D.K.\ gratefully acknowledges funding from the DFG through an Emmy Noether Research Group (grant number KR4801/1-1).
J.M.D.K.\ and S.T.G.\ gratefully acknowledge funding from the European Research Council (ERC) under the European Union's Horizon 2020 research and innovation programme via the ERC Starting Grant MUSTANG (grant agreement number 714907).
COOL Research DAO is a Decentralised Autonomous Organisation supporting research in astrophysics aimed at uncovering our cosmic origins.
S.T.G.\ gratefully acknowledges the generous and invaluable support of the Klaus Tschira Foundation.
J.L.P.\ is supported by the Australian government through the Australian Research Council's Discovery Projects funding scheme (DP220101863).
R.A.C.\ is a Royal Society University Research Fellow.
M.R.C.\ gratefully acknowledges the Canadian Institute for Theoretical Astrophysics (CITA) National Fellowship for partial support; this work was supported by the Natural Sciences and Engineering Research Council of Canada (NSERC).
P.S.J.\ would like to thank Anna Pasquali for feedback and helpful discussions.
This study made use of the Prospero high performance computing facility at Liverpool John Moores University. \\

\noindent\textit{Software:} {
\package{matplotlib} \citep{Hunter_2007},
\package{numpy} \citep{vanderWalt_2011}.
}\\

\bibliographystyle{aasjournal}
\bibliography{references}

\begin{appendix}
\section{Impact of Upper Metallicity Cuts}
\label{appendix}

\begin{figure}
    \centering
    \includegraphics[width=0.55\columnwidth]{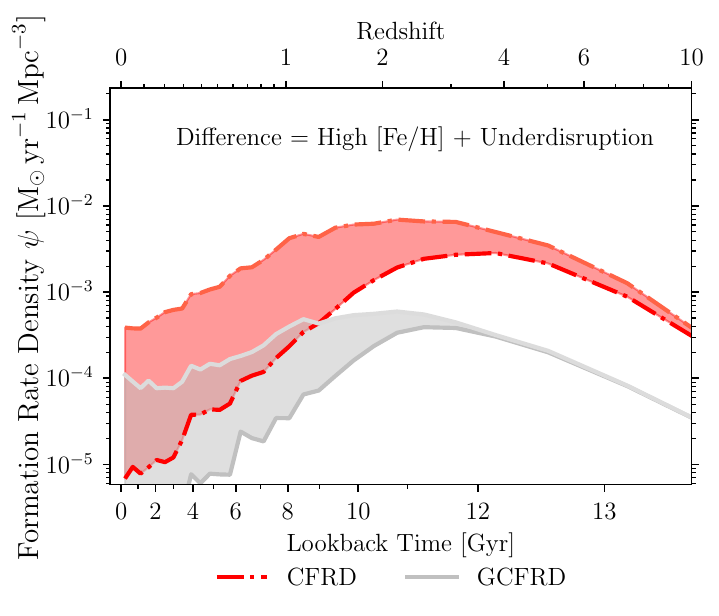}
    \caption{Difference in the formation rate densities of clusters (\textit{red}) and GCs (\textit{grey}) caused by the application of the upper metallicity limits for clusters and GCs given in Table~\ref{tab:metallicity_range}. The lighter lines track the formation rate densities without upper metallicity cuts, while the darker ones follow the formation rate densities with the upper [Fe/H] limit. The difference only becomes significant in predicting the GC formation rate densities at redshifts $z \lesssim 2$, i.e.\ at times more recent than the cosmic epoch that the predictions of this work mostly focus on. As a result, our conclusions are unaffected by the metallicity limits.}
    \label{fig:difference_metcuts}
\end{figure}

In this paper, we apply upper limits to the metallicities of clusters and GCs, as described in Section~\ref{sec:E-MOSAICS}. \citet{Pfeffer+2023} show that the metallicity distribution generally agrees with observations except for the range $10.5 < \log(M_{\star,\rm gal} / \MSun) < 11$. However, we cannot distinguish between high-metallicity clusters and GCs that should survive and those that are only retained in the simulation due to underdisruption (see appendix~D of \citealt{Kruijssen_2019a} for details on how this affects high-metallicity GCs). Therefore, we illustrate in Figure~\ref{fig:difference_metcuts} how the upper [Fe/H] limits on the cluster and GC population (see Table~\ref{tab:metallicity_range}) impact the formation histories. The figure shows how the CFRD and GCFRD from Fig.~\ref{fig:formhist_densities} change by removing the upper metallicity limits. We see that the application of these limits has a significant impact on the formation rate densities, but only affects the GCs after $z\lesssim 2$, with a difference of more than an order of magnitude at $z=0$ for both CFRD and GCFRD. This difference clearly illustrates the need for these limits, as documented extensively in earlier papers by the E-MOSAICs team. By including these metallicity limits, we improve the predictive strength of our results and provide a better comparability with future observations.
\end{appendix}

\end{document}